# Topological Fano Resonances

Farzad Zangeneh-Nejad and Romain Fleury*

*Laboratory of Wave Engineering, School of Electrical Engineering, EPFL, Station 11, 1015 Lausanne, Switzerland*

*To whom correspondence should be addressed. Email: romain.fleury@epfl.ch

**The Fano resonance is a widespread wave scattering phenomenon associated with a peculiar asymmetric and ultra-sharp line shape, which has found applications in a large variety of prominent optical devices. While its substantial sensitivity to geometrical and environmental changes makes it the cornerstone of efficient sensors, it also renders the practical realization of Fano-based systems extremely challenging. Here, we introduce the concept of topological Fano resonance, whose ultra-sharp asymmetric line shape is guaranteed by design and protected against geometrical imperfections, yet remaining sensitive to external parameters. We report the experimental observation of such resonances in an acoustic system, and demonstrate their inherent robustness to geometrical disorder. Such topologically-protected Fano resonances, which can also be found in microwave, optical and plasmonic systems, open up exciting frontiers for the generation of various reliable wave-based devices including low-threshold lasers, perfect absorbers, ultrafast switches or modulators, and highly accurate interferometers, by circumventing the performance degradations caused by inadvertent fabrication flaws.**

The Fano resonance [1] is a ubiquitous scattering wave phenomenon, commonly found in various branches of science and engineering like atomic and solid states physics [2,3], electromagnetism [4-6], electronic circuits [7,8], photonics [9-15], nonlinear optics [16,17], and acoustics [18,19]. Such intriguing type of resonance occurs as a result of constructive and destructive interferences between two overlapping resonant scattering states with different lifetimes: a wide-band "bright" resonance, and a narrower "dark" resonance. The dark resonance serves as a quasi-isolated (localized) state whose coexistence in the continuum of the bright resonance gives rise to the asymmetric and ultra-sharp line shape profile of the Fano resonance [20].

While originally developed to explain the inelastic scattering of electrons in Helium [21], the ultra-sharp spectrum of Fano resonances has established itself as a centerpiece in the realization of many modern optical devices, including low threshold lasers [22], low energy switches [23], ultrafast modulators [24], high quality factor filters [25], compact electromagnetically induced transparency (EIT) devices [26], ultrathin perfect absorbers [27], and highly accurate interferometers [28]. Moreover, apart from its steepness, the peculiar asymmetric line shape of Fano resonances is found to be excessively sensitive to environmental changes, a characteristic which has enabled the realization of highly sensitive and accurate sensors [29,30].

Unfortunately, this extreme sensitivity also comes with a price: guaranteeing a Fano resonance near the frequency of interest requires extreme control over the system's geometry as fabrication imperfections can shift the bright and dark modes away from each other. Even worse, disorder may introduce extra parasitic resonances, completely destroying the much-sought Fano line shape. Therefore, the performance advantages obtained from Fano interferences are often mitigated by the costs associated with the required fabrication technology. Although they work quite well in theory, the performance of Fano-based devices degrades significantly upon fabrication due to inevitable flaws [31].

The recent development of topological wave physics [32-34], which has demonstrated a wide class of topologically-protected photonic [35-43], mechanical [44-47] or acoustic states [48-57], may offer an unprecedented solution to this vexing problem. It has been demonstrated that, depending on the characteristics of the topology of the bulk bands, topological edge modes can be protected to certain classes of defects [58-64]. Such protection has inspired the realization of a new generation of topological devices, such as lasers and modulators, with unprecedented robustness to fabrication flaws [65,66]. Following these pioneering works, one may wonder whether topology may be leveraged to build a novel form of sturdy topologically-protected Fano resonance, which may be largely appealing for a wide range of applications in different areas of physics, from photonics to mechanics and acoustics.

In this Letter, we extend the reach of topological wave physics by introducing the general concept of topological Fano resonances. We demonstrate that the ultra-sharp spectrum of Fano resonances can be guaranteed by design without stringent geometrical requirements, and with a complete immunity to structural disorder, while retaining its ability to shift under environmental changes and act as a sensor. We further provide an experimental demonstration of topological Fano resonances in a realistic acoustic system.

The underlying idea of our proposal is conceptually sketched in Fig. 1. Let us first consider the conventional Fano resonance (Fig. 1a). It is typically achieved by overlapping two resonant states with different lifetimes: one needs a bright and a dark mode that coexist in a certain spectral range. Upon coupling, the resonant interaction of these two states leads to constructive and destructive interferences, which creates the archetypal asymmetric peak-and-dip Fano line shape [67]. Since the Fano resonance is a resonant scattering state obtained from an interference phenomenon between only two modes, it is intrinsically sensitive to perturbations in the geometry and environment. More specifically, even small amount of disorder can shift the resonance frequencies and coupling phase of the two resonating modes, and introduce new resonant states that uncontrollably deform the peculiar Fano spectral signature. This is schematically represented in the right panel of Fig. 1a, where disorder has created new dips and peaks in the spectrum. Our proposal is, instead, to start from bright and dark modes whose existence is topologically guaranteed (Fig. 1b, left). Upon coupling, a topological Fano resonance may be created, whose line shape can be preserved even in the presence of geometrical disorder. Environmental changes may shift the bright and dark modes, but not suppress them nor introduce new modes. As a result, the Fano shape inherits some form of topological immunity against disorder, as we demonstrate in the following.

Following our initial vision, we need a topological dark mode to interact with a topological bright mode. Here, we achieve this in an acoustic system (see the supplementary material [68] for an electromagnetic equivalent), by creating two topological subspaces based on the physics of symmetry-protected bound states in the continuum. Consider a two-dimensional acoustic parallel-plate waveguide with a plate separation of *2h=10 cm*, containing cylindrical obstacles placed at the centerline and arranged in a periodic lattice with lattice constant of *a* (Fig. 2a, top). The corresponding band structure of the crystal is shown in the bottom panel. It is instructive to divide the dispersion bands into two categories according to the symmetry of their corresponding mode profiles, which are represented in Fig. 2b for $k_x = 0$ and $k_x = \pm \pi/a$. The first category is associated with eigenmodes having even symmetry with respect to the centerline of the waveguide. These bands, marked in blue in Fig. 2a, possess the typical frequency dispersion of a one-dimensional sonic crystal made of far-field coupled scatterers, with waves propagating down to the quasi-static limit. The second category (the red band), corresponds to odd-symmetric eigenstates, exhibiting the typical cosine band characteristic of evanescently coupled resonators. Such odd-symmetric modes originate from symmetry-protected bound states in the continuum (BICs) [69-72], which are prevented to leak into the continuum of even-symmetric waveguide modes (see [68], section I). Interestingly, the odd band behaves differently than the even bands upon scaling: while shrinking the lattice constant shifts all the even bands up in frequency, the position of the BIC band stays constant [68]. This allows us to overlap the second blue band and the first red band in Fig. 1a (setting *a=16.3 cm*), creating two uncoupled physical subspaces whose topological properties can be further simultaneously engineered, as we now demonstrate.

To induce a non-trivial topology in such one-dimensional periodic structure, we first double the size of the unit cell, now considering the lattice as a repetition of cylinders pairs, which effectively folds the band structure. Next, we lift the degeneracies of the new band structure at the folding points by reducing or increasing the distance between the two cylinders within the new extended unit cell. The top insets of Figs. 2c and d illustrate the geometries of the obtained shrunk and expanded lattices, respectively. The bottom insets report the corresponding band structures, where the bands are colored according to the previously explained symmetry classification. Similar to a SSH chain [73], while both configurations exhibit exactly the same band structure, they are topologically distinct, as confirmed by Zak phase calculations [68], whose results are summarized on the right-hand sides of the dispersion diagrams.

To form the topological Fano resonance, we now consider the topological edge states that form at an interface between these two crystals around 1.5 kHz (by design, the topological band-gaps of the even and odd subspaces overlap around this frequency). In Fig. 3a, we form a finite-size system made of four crystal cells on each side of the interface (top inset). For the sake of simplicity, we have removed the last obstructing cylinder from the non-trivial sample so as to avoid an extra BIC topological edge mode to form at the right hand side interface with the empty waveguide (which is a trivial insulator in the odd-symmetric subspace). Removing this last cylinder does not fundamentally change our results, but creates a single Fano-resonance instead of a double one. Similar to our previous classification, the obtained edge states can be categorized into two classes: the dark edge state stemming from the BIC mode, distinguished by its odd profile, and the bright one originating from the even modes (see [68], section IV). The two-port system is then excited by a plane wave incident from the left and the transmission coefficient is extracted (bottom plot in Fig. 3a). As observed, the spectrum shows a single peak right at the resonance frequency of the bright edge mode (the red dashed line), because the incident plane wave cannot excite the odd-symmetric (dark) edge mode. To let, instead, the bright and dark modes interact and create a topological Fano resonance, we must break the vertical inversion symmetry and slightly move all cylinders up (or down) from the center line. Doing so indeed yields the expected Fano line shape (the solid blue line). It should be emphasized that the slight movement of obstacles from the centerline does not change the topological classification of the bands (see [68]; section III). Consequently, both of the obtained bright and dark resonances are of topological nature, and the origin of the Fano resonance is indeed rooted to the topological properties of the surrounding bulk insulators. Therefore, we expect the presence of the Fano resonance to be guaranteed even in the presence of disorder, as long as it is not strong enough to close the band gaps. To test this hypothesis, we randomly change the positions of the obstructing circles in Fig. 3b (1.5 cm average shift) and repeat the scattering experiment. The transmission spectrum of the waveguide, represented in Fig. 3b (bottom panel), reveals three important properties. Despite the large degree of disorder, (i) the Fano-resonance is still present, (ii) the Fano shape is not disturbed by any new localized-mode (which would add new peaks/dips) (iii) the Fano resonance peak and dip can shift and is still sensitive to environmental changes. These features remain true for any realization of disorder. This is illustrated in Fig. 3c, which reports the evolution of the transmission spectrum versus disorder strength. This plot confirms that not only disorder slightly changes the resonance frequencies of the bright and dark edge modes, but also it does not affect the presence and representative peak-and-dip shape of the Fano resonance. Next, we compare the robustness of the topological Fano to the one of a topologically trivial Fano resonance, like the one considered in Fig. 3d, which is based on bright and dark resonant defect-tunneling through a Bragg band gap [74]. In this case, the trivial Fano spectrum is deeply affected by the presence of disorder, which induces new localized states that destroy its characteristic shape (Fig. 3e). Comparing the parametric plot of Fig. 3f to the one of Fig. 3c, we see that the trivial Fano resonance only survives low disorder levels, until is completely destroyed by a disorder-induced localized mode.

Based on these findings we have built a prototype to demonstrate experimentally the robustness of the topological Fano resonance (Fig. 4a, top panel). The sample consists of an acrylic extruded clear square tube ($\rho = 1.18 \frac{g}{cm^3}$ and *B=3300 MPa*) serving as the acoustic waveguide (cross section is *49 cm²*), and nylon black rods ($\rho = 1.15 \frac{g}{cm^3}$ and *B=3400 MPa*) with diameters of *3.5 cm*. We then consider a scattering experiment similar to that of Fig. 3a. Numerical finite-elements simulations including realistic visco-thermal losses predict a topological Fano resonance around the frequency *$f_0$=2.3 kHz*. To confirm this prediction, we sent burst noise from the left of the system, whereas the right-hand side port was connected to a broadband anechoic termination (see [68], section VII). We then measured the corresponding transmission coefficient, which is represented in the bottom panel of Fig. 4a (blue line) and compared to the simulation results (dashed red line). As expected, we observe the topological Fano transmission spectrum (circled in green), with very good agreement with simulations. Next, we randomly moved the obstacles from their real positions to get the disordered configuration shown in the top panel of Fig. 4b. The measured transmission coefficient, represented in the bottom panel, demonstrates that the Fano line shape is indeed perfectly preserved, despite the large level of disorder (1cm average shift, i.e. 14% of the waveguide width, with no preferred direction). To confirm the topological origin of this property, we repeated the scattering experiment for a (trivial) Bragg-induced Fano resonance obtained from a periodic lattice with a missing center rod at the center (total of six rods, slightly up-shifted with respect to the center line, see Fig. 4c). Numerical simulations show that the missing rod creates an even defect mode overlapping with an odd dark state carried by the Bragg structure, i.e. a trivial Fano resonance. The measured spectrum exactly confirms these predictions. By adding, in Fig. 4d, the same level of disorder as in Fig 4b (1 cm average disorder strength, with no preferred direction), it is clear that the initial Fano shape of the resonance is extremely affected: the characteristic peak-and-dip shape of the Fano is completely destroyed, and a new dip is added to the spectrum. Such drastic deformations would be highly detrimental in any sensing application, which are usually based on tracking the shifts of the Fano dip.

To conclude, we have introduced and demonstrated the concept of topological Fano resonance. Our theoretical and experimental findings demonstrate the superior robustness of topological Fano resonances over trivial Fano resonances. Topology forces the archetypal Fano line shape to occur in the desired frequency range, as long as the level of disorder is not strong enough to close the surrounding topological band gaps. In addition, topological Fano resonances are still sensitive to environmental changes, and tracking the shift of the Fano dip may allow for a new generation of sturdy sensors. Our experiment shows that obtaining a topological Fano resonance does not require tight geometrical tolerances, very different from trivial Fano resonances. We believe that the concept of topological Fano resonance can offer new perspectives in many applicative fields, including phononic, photonic and plasmonic sensing technologies, all-optical information processing, or biomolecular detection. In addition, the reliability of topological Fano resonances offers promising perspectives for the realization of various fabrication-insensitive optical devices such as low-threshold lasers, ultrafast modulators and switches, and perfect absorbers.

# Figures

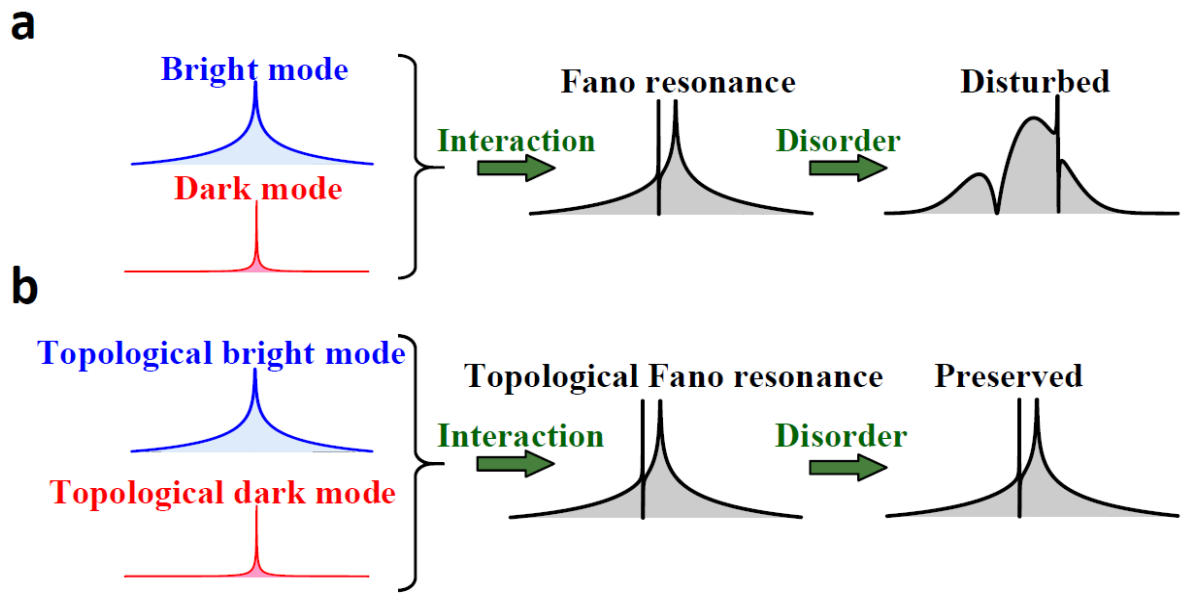

**Fig. 1: Topological Fano resonance**. **a,** Interaction between a bright resonance and a dark narrower resonance can lead to an ultra-sharp and asymmetric line shape, characteristic of a Fano resonance. Even small levels of disorder, however, can severely destroy the line shape of the resonance, by introducing new dips and peaks. **b,** To make Fano resonances immune to disorder, one can instead start from bright and dark modes whose existence is topologically guaranteed. The resulting topological Fano line shape is robust against a large class of geometrical imperfections: the occurrence of new disorder-induced dips and peaks is prevented by topology.

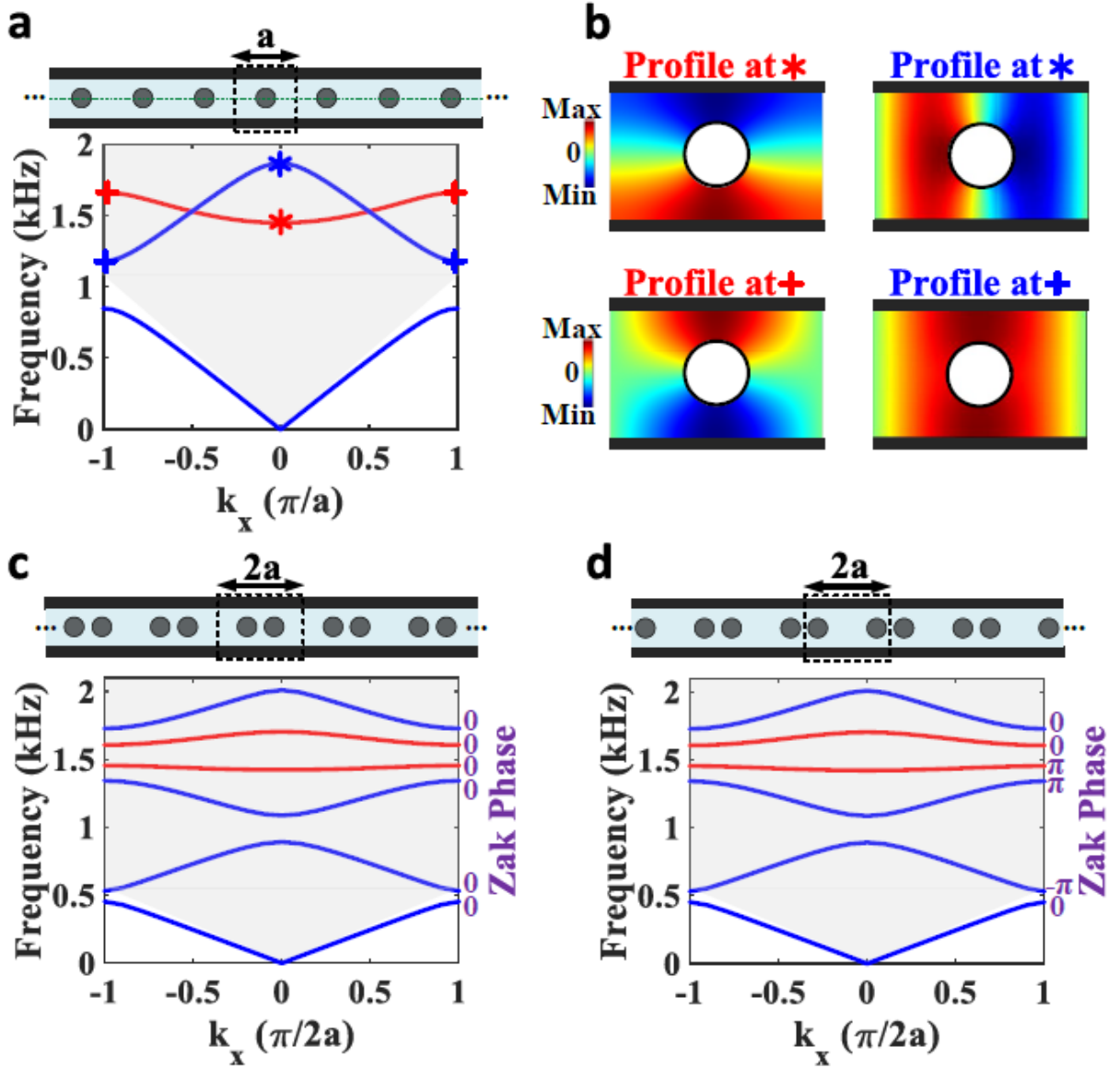

**Fig. 2: Independent topological subspaces in an acoustic waveguide. a,** Band structure of an acoustic parallel plate waveguide containing obstructing cylinders placed on its center line, arranged in a periodic lattice. The red band is the dispersion of an odd-symmetric eigenmode (originating from evanescently-coupled symmetry-protected bound states in the continuum), while the blue bands correspond to regular phononic crystal bands with even mode symmetry. The grey area represents the empty waveguide continuum. **b,** Profiles of the odd and even modes at specific Bloch wavenumbers. **c,** Band structure of the crystal when considering the extended unit cell (which includes two periods), and reducing the distance between the two obstacles with respect to the folded case. **d**, Same as panel c except that the distance between the obstacles is increased. Calculations of the Zak phases of the bands (purple) demonstrate that panels c and d correspond to distinct topological phases. The band gaps of the two topological subspaces overlap around 1.5 kHz.

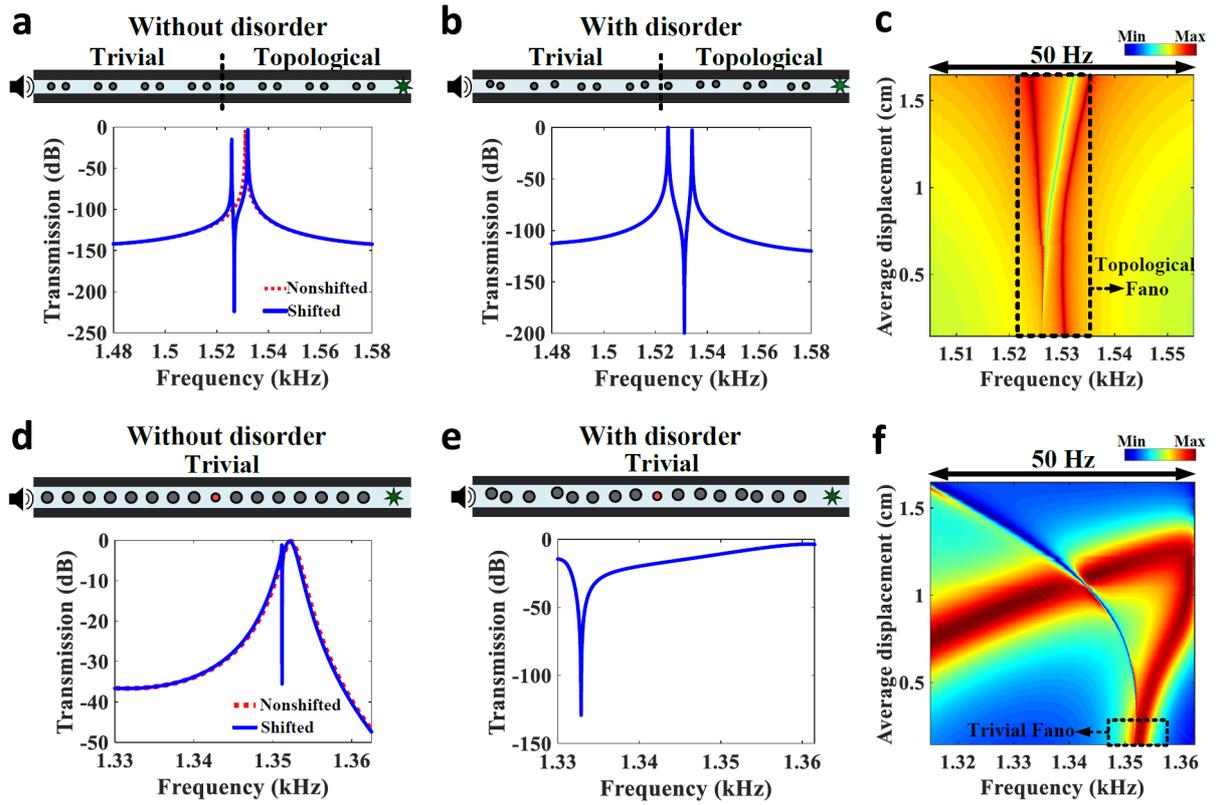

**Fig. 3: Full-wave numerical demonstration of topologically-protected Fano resonances, a,** Four unit cells from the trivial lattice of Fig. 2c are connected to four cells from the non-trivial system of Fig. 2d. Even and odd topological edge modes accordingly form at the interface between the two insulators. By sending a plane wave from the left, however, only the even edge mode can be excited (the red dashed line), leading to only one resonance in the transmission spectrum. This is no longer the case if the obstacles are slightly shifted away from the centerline, allowing even and odd modes to interact, and inducing a topological Fano resonance (the solid blue line). **b,** Transmission spectrum of the waveguide when the obstacles are randomly moved from their original places. The Fano line shape is preserved due to topology. **c,** Evolution of the transmission spectrum versus disorder strength, **d,e,f,** Same as (a,b,c) for a similar acoustic system with a topologically trivial Fano resonance. Unlike the topological Fano resonance, the trivial Fano resonance only survives very low disorder levels.

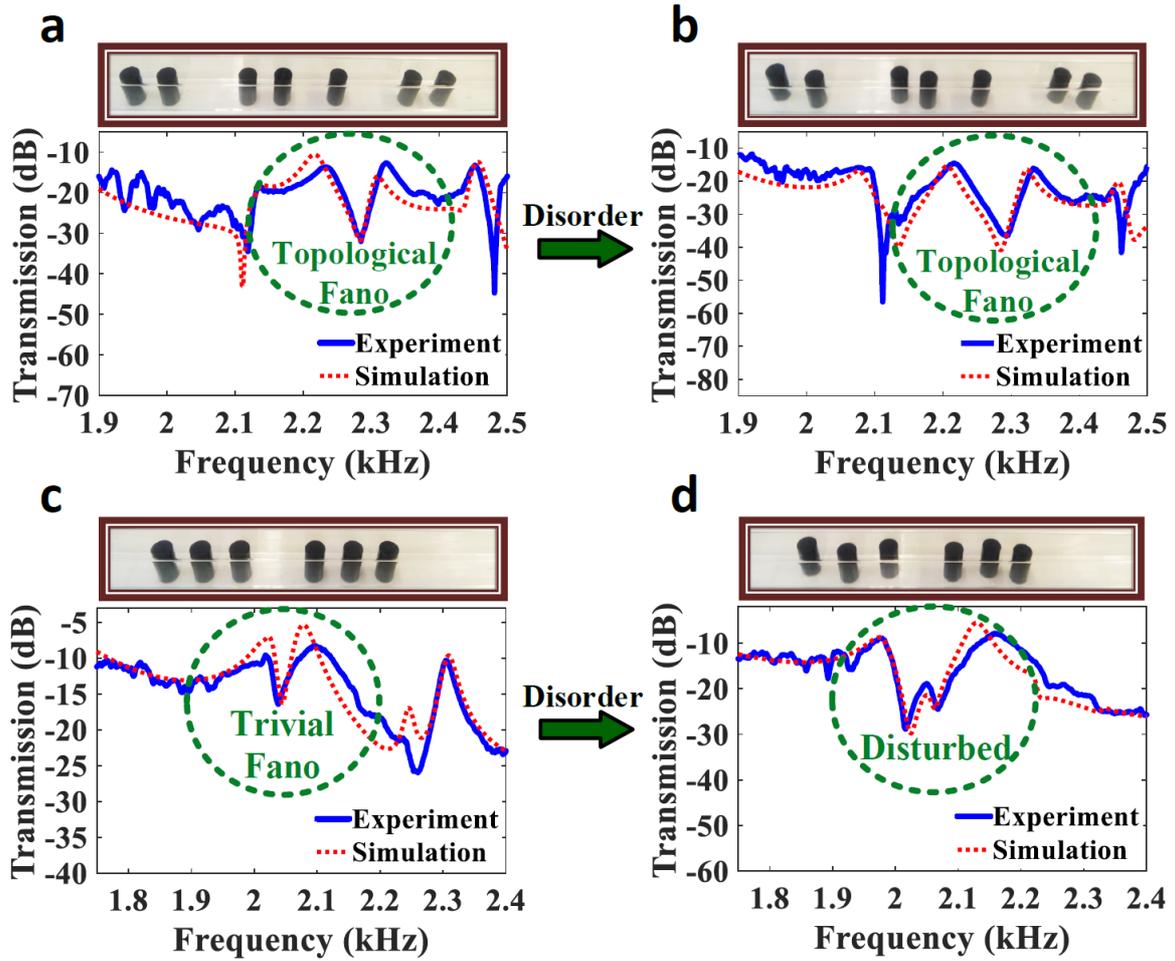

**Fig. 4: Experimental validation of topological Fano resonances and their robustness, a,** Nylon black rods are embedded inside a transparent square acoustic waveguide, implementing a scattering experiment analogous to Fig. 3a. The structure supports a topological Fano resonance around the frequency $f_0=2.3\ kHz$ as observed in the bottom panel. **b,** The obstructing rods are randomly moved away from their original positions, introducing position disorder. The Fano lineshape is maintained. **c,d,** Same as (a,b) for a trivial Fano resonance induced when coupling a topologically trivial Bragg defect mode and a BIC. The initial Fano lineshape is destroyed when applying disorder of the same strength as in panel b.

## Acknowledgments

This work was supported by the Swiss National Science Foundation (SNSF) under Grant No. 172487.

# References

1. U. Fano, Phys. Rev. **124**, 1866 (1961).
2. O. Újsághy, J. Kroha, L. Szunyogh, and A. Zawadowski, Phys. Rev. Lett. **85**, 2557 (2000).
3. A. C. Johnson, C. M. Marcus, M. P. Hanson, and A. C. Gossard, Phys. Rev. Lett. **93**, (2004).
4. S. Rotter, F. Libisch, J. Burgdörfer, U. Kuhl, and H. J. Stöckmann, Phys. Rev. E - Stat. Physics, Plasmas, Fluids, Relat. Interdiscip. Top. **69**, 4 (2004).
5. A. Bärnthaler, S. Rotter, F. Libisch, J. Burgdörfer, S. Gehler, U. Kuhl, and H. J. Stöckmann, Phys. Rev. Lett. **105**, (2010).
6. E. O. Kamenetskii, G. Vaisman, and R. Shavit, J. Appl. Phys. **114**, (2013).
7. A. Attaran, S. D. Emami, M. R. K. Soltanian, R. Penny, F. behbahani, S. W. Harun, H. Ahmad, H. A. Abdul-Rashid, and M. Moghavvemi, Plasmonics **9**, 1303 (2014).
8. B. Lv, R. Li, J. Fu, Q. Wu, K. Zhang, W. Chen, Z. Wang, and R. Ma, Sci. Rep. **6**, (2016).
9. B. Luk'Yanchuk, N. I. Zheludev, S. A. Maier, N. J. Halas, P. Nordlander, H. Giessen, and C. T. Chong, Nat. Mater. **9**, 707 (2010).
10. F. Shafiei, F. Monticone, K. Q. Le, X. X. Liu, T. Hartsfield, A. Alù, and X. Li, Nat. Nanotechnol. **8**, 95 (2013).
11. C. Wu, A. B. Khanikaev, and G. Shvets, Phys. Rev. Lett. **106**, (2011).
12. P. Fan, Z. Yu, S. Fan, and M. L. Brongersma, Nat. Mater. **13**, 471 (2014).
13. C. Argyropoulos, F. Monticone, G. Daguanno, and A. Alù, Appl. Phys. Lett. **103**, (2013).
14. M. V. Rybin, A. B. Khanikaev, M. Inoue, K. B. Samusev, M. J. Steel, G. Yushin, and M. F. Limonov, Phys. Rev. Lett. **103**, (2009).
15. J. B. Lassiter, H. Sobhani, J. A. Fan, J. Kundu, F. Capasso, P. Nordlander, and N. J. Halas, Nano Lett. **10**, 3184 (2010).
16. Y. Wang, L. Liao, T. Hu, S. Luo, L. Wu, J. Wang, Z. Zhang, W. Xie, L. Sun, A. V. Kavokin, X. Shen, and Z. Chen, Phys. Rev. Lett. **118**, (2017).
17. J. Butet and O. J. F. Martin, Opt. Express **22**, 29693 (2014).
18. E. H. El Boudouti, T. Mrabti, H. Al-Wahsh, B. Djafari-Rouhani, A. Akjouj, and L. Dobrzynski, J. Phys. Condens. Matter **20**, (2008).
19. S. Hein, W. Koch, and L. Nannen, J. Fluid Mech. **664**, 238 (2010).
20. M. F. Limonov, M. V. Rybin, A. N. Poddubny, and Y. S. Kivshar, Nat. Photonics **11**, 543 (2017).
21. R. P. Madden, and K. Codling, The Astrophysical Journal, **141**, 364 (1965).
22. S.L. Chua, Y. Chong, A. D. Stone, M. Soljacic, and J. Bravo-Abad, Opt. Express **19**, 1539 (2011).
23. K. Nozaki, A. Shinya, S. Matsuo, T. Sato, E. Kuramochi, and M. Notomi, Opt. Express **21**, 11877 (2013).
24. W. Zhao, H. Jiang, B. Liu, Y. Jiang, C. Tang, and J. Li, Appl. Phys. Lett. **107**, (2015).
25. W. Cao, R. Singh, I. A. I. Al-Naib, M. He, A. J. Taylor, and W. Zhang, Opt. Lett. **37**, 3366 (2012).
26. N. Papasimakis and N. I. Zheludev, Opt. Photonics News **20**, 22 (2009).
27. K. Q. Le and J. Bai, J. Opt. Soc. Am. B **32**, 595 (2015).
28. K. P. Heeg, C. Ott, D. Schumacher, H. C. Wille, R. Röhlsberger, T. Pfeifer, and J. Evers, Phys. Rev. Lett. **114**, (2015).

29. C. Wu, A. B. Khanikaev, R. Adato, N. Arju, A. A. Yanik, H. Altug, and G. Shvets, Nat. Mater. **11**, 69 (2012).
30. Z. Li, S. Zhang, L. Tong, P. Wang, B. Dong, and H. Xu, ACS Nano **8**, 701 (2014).
31. Y. Yu, M. Heuck, H. Hu, W. Xue, C. Peucheret, Y. Chen, L. K. Oxenløwe, K. Yvind, and J. Mørk, Appl. Phys. Lett. **105**, (2014).
32. C. L. Kane and E. J. Mele, Phys. Rev. Lett. **95**, (2005).
33. M. Fruchart and D. Carpentier, Comptes Rendus Phys. **14**, 779 (2013).
34. F. D. M. Haldane, Phys. Rev. Lett. **61**, 2015 (1988).
35. L. Lu, J. D. Joannopoulos, and M. Soljačić, Nat. Photonics **8**, 821 (2014).
36. W. J. Chen, S. J. Jiang, X. D. Chen, B. Zhu, L. Zhou, J. W. Dong, and C. T. Chan, Nat. Commun. **5**, (2014).
37. X. Cheng, C. Jouvaud, X. Ni, S. H. Mousavi, A. Z. Genack, and A. B. Khanikaev, Nat. Mater. **15**, 542 (2016).
38. M. C. Rechtsman, J. M. Zeuner, Y. Plotnik, Y. Lumer, D. Podolsky, F. Dreisow, S. Nolte, M. Segev, and A. Szameit, Nature **496**, 196 (2013).
39. T. Ma, A. B. Khanikaev, S. H. Mousavi, and G. Shvets, Phys. Rev. Lett. **114**, (2015).
40. X. Cheng, C. Jouvaud, X. Ni, S. H. Mousavi, A. Z. Genack, and A. B. Khanikaev, Nat. Mater. **15**, 542 (2016).
41. L. J. Maczewsky, J. M. Zeuner, S. Nolte, and A. Szameit, Nat. Commun. **8**, (2017).
42. M. Hafezi, S. Mittal, J. Fan, A. Migdall, and J. M. Taylor, Nat. Photonics **7**, 1001 (2013).
43. T. Kitagawa, M. A. Broome, A. Fedrizzi, M. S. Rudner, E. Berg, I. Kassal, A. Aspuru-Guzik, E. Demler, and A. G. White, Nat. Commun. **3**, (2012).
44. S. D. Huber, Nat. Phys. **12**, 621 (2016).
45. R. Chaunsali, E. Kim, A. Thakkar, P. G. Kevrekidis, and J. Yang, Phys. Rev. Lett. **119**, (2017).
46. B. G. G. Chen, B. Liu, A. A. Evans, J. Paulose, I. Cohen, V. Vitelli, and C. D. Santangelo, Phys. Rev. Lett. **116**, (2016).
47. S. Roman and D. H. Sebastian, Science. **349**, 47 (2015).
48. Z. Yang, F. Gao, X. Shi, X. Lin, Z. Gao, Y. Chong, and B. Zhang, Phys. Rev. Lett. **114**, (2015).
49. M. Xiao, G. Ma, Z. Yang, P. Sheng, Z. Q. Zhang, and C. T. Chan, Nat. Phys. **11**, 240 (2015).
50. R. Fleury, A. B. Khanikaev, and A. Alù, Nat. Commun. **7**, (2016).
51. A. B. Khanikaev, R. Fleury, S. H. Mousavi, and A. Alù, Nat. Commun. **6**, (2015).
52. C. He, X. Ni, H. Ge, X. C. Sun, Y. Bin Chen, M. H. Lu, X. P. Liu, and Y. F. Chen, Nat. Phys. **12**, 1124 (2016).
53. Z. Zhang, Q. Wei, Y. Cheng, T. Zhang, D. Wu, and X. Liu, Phys. Rev. Lett. **118**, (2017).
54. J. Mei, Z. Chen, and Y. Wu, Sci. Rep. **6**, (2016).
55. J. Lu, C. Qiu, L. Ye, X. Fan, M. Ke, F. Zhang, and Z. Liu, Nat. Phys. **13**, 369 (2017).
56. V. Peano, C. Brendel, M. Schmidt, and F. Marquardt, Phys. Rev. X **5**, (2015).
57. X. Ni, C. He, X. C. Sun, X. P. Liu, M. H. Lu, L. Feng, and Y. F. Chen, New J. Phys. **17**, (2015).
58. S. Yves, R. Fleury, T. Berthelot, M. Fink, F. Lemoult, and G. Lerosey, Nat. Commun. **8**, (2017).
59. P. Wang, L. Lu, and K. Bertoldi, Phys. Rev. Lett. **115**, (2015).
60. M. Hafezi, E. A. Demler, M. D. Lukin, and J. M. Taylor, Nat. Phys. **7**, 907 (2011).
61. Z. Wang, Y. Chong, J. D. Joannopoulos, and M. Soljačić, Nature **461**, 772 (2009).
62. W. Gao, M. Lawrence, B. Yang, F. Liu, F. Fang, B. Béri, J. Li, and S. Zhang, Phys. Rev. Lett. **114**, (2015).

63. Y. Ran, Y. Zhang, and A. Vishwanath, Nat. Phys. **5**, 298 (2009).
64. C. He, Z. Li, X. Ni, X. C. Sun, S. Y. Yu, M. H. Lu, X. P. Liu, and Y. F. Chen, Appl. Phys. Lett. **108**, (2016).
65. G. Harari, M. A. Bandres, Y. Lumer, M. C. Rechtsman, Y. D. Chong, M. Khajavikhan, D. N. Christodoulides, and M. Segev, Science. **359**, (2018).
66. H. Yu, H. Zhang, Y. Wang, C. Zhao, B. Wang, S. Wen, H. Zhang, and J. Wang, Laser Photonics Rev. **7**, (2013).
67. A. E. Miroshnichenko, S. Flach, and Y. S. Kivshar, Rev. Mod. Phys. **82**, 2257 (2010).
68. F.Zangeneh-Nejad, R. Fleury, Topological Fano resonances, Supplementary materials.
69. Y. X. Xiao, G. Ma, Z. Q. Zhang, and C. T. Chan, Phys. Rev. Lett. **118**, (2017).
70. M. Rybin and Y. Kivshar, Nature **541**, 164 (2017).
71. S. T. Ha, Y. H. Fu, N. K. Emani, Z. Pan, R. M. Bakker, R. Paniagua-Domínguez, and A. I. Kuznetsov, Nat. Nanotechnol. (2018).
72. C. W. Hsu, B. Zhen, A. D. Stone, J. D. Joannopoulos, and M. Soljacic, Nat. Rev. Mater. **1**, (2016).
73. J. K. Asbóth, L. Oroszlány, and A. Pályi, Lect. Notes Phys. **919**, (2015).
74. O. Painter, Science. **284**, 1819 (1999).